\documentclass{emulateapj}
\RequirePackage{lineno}
\usepackage{amsmath}
\usepackage{amssymb}
\usepackage{amsthm}
\usepackage{array}
\usepackage{float}
\usepackage{graphicx}
\usepackage{subfigure}
\usepackage{color}

\newcounter{ichi}
\newcounter{ni}
\newcounter{san}
\newcounter{yon}
\slugcomment{}

\shorttitle{Electromagnetic counterparts of black hole mergers}
\shortauthors{Murase et al.}

\begin{document}

\title{Ultrafast Outflows from Black Hole Mergers with A Mini-Disk}
\author{Kohta Murase\altaffilmark{1}, Kazumi Kashiyama\altaffilmark{2}, Peter M\'esz\'aros\altaffilmark{1}, Ian Shoemaker\altaffilmark{1}, and Nicholas Senno\altaffilmark{1}}
\altaffiltext{1}{Center for Particle and Gravitational Astrophysics; Department of Physics; Department of Astronomy \& Astrophysics, The Pennsylvania State University, University Park, PA 16802, USA}
\altaffiltext{2}{Einstein Fellow -- Theoretical Astrophysics Center, Department of Astronomy, University of California, Berkeley, CA 94720, USA}


\begin{abstract}
Recently, the direct detection of gravitational waves from black hole (BH) mergers was announced by the 
Advanced LIGO Collaboration.  Multi-messenger counterparts of stellar-mass BH mergers are of interest, 
and it had been suggested that a small disk or celestial body may be involved in the binary of two BHs. 
To test such possibilities, we consider the fate of a wind powered by an active mini-disk in a relatively short, super-Eddington accretion 
episode onto a BH with $\sim10-100$ solar masses.
We show that its thermal emission could be seen as a fast optical transient with the duration from hours to days.  
We also find that the coasting outflow forms external shocks due to interaction with the interstellar 
medium, whose synchrotron emission might be expected in the radio band on a time scale of years.  
Finally, we also discuss a possible jet component and the associated high-energy neutrino emission as well as ultra-high-energy cosmic-ray acceleration. 
\end{abstract}


\section{Introduction}
The discovery of gravitational waves (GWs) from GW 150914 by Advanced LIGO opens a new 
window of the high-energy universe~\citep{GW15094_LIGO_Main}.  
This first GW detection has simultaneously yielded the first observation of a binary black hole (BH) merger.  
The inferred initial masses $36^{+5}_{-4}M_{\odot}$ and $29^{+4}_{-4}M_{\odot}$ merged to form a final black 
hole of mass $62^{+4}_{-4}M_{\odot}$, with the difference in final and initial masses corresponding to the 
energy emitted in GW radiation~\citep{GW15094_LIGO_Main}.  
This marks the beginning of gravitational wave astronomy~\citep{2016ApJ...818L..22A}, 
and offers a completely orthogonal means of observing the cosmos compared to the traditional avenues 
afforded by electromagnetically-based telescopes. 

It also heralds the beginning of a new era in multi-messenger astrophysics, in which both electromagnetic (EM), neutrino, and GW probes are combined.  
For any class of GW sources, identifying EM and/or neutrino counterparts of the GW sources will enable us not only to study dynamics and emission 
mechanisms of the transients but also to obtain clues to environments where the sources are formed.   
EM-based telescopes have better localization capabilities than GW detectors, and strategic searches have been anticipated especially
for binary systems involving a neutron star (NS) such as NS-NS and NS-BH mergers~\citep[e.g.,][]{2013ApJ...767..124N}.  
On the other hand, an obvious EM counterpart is unexpected from BH-BH binary mergers in vacuum. 
As discussed in the context of super-massive black hole binaries in the nucleus~\citep{2012MNRAS.423L..65B}, expected EM signals depend on details of the setup.    
In other words, any counterpart would reveal the non-trivial presence of matter around BHs.
For example, long-lived disks with small masses may be formed around BHs, which may lead to the possible existence of planets orbiting BHs~\citep{Perna_et_al_2014}.
Also, it has been suggested that a BH binary in a hierarchal three-body system may resonantly trap a star and its tidal disruption might emit EM signals around the 
coalescence of the two BHs~\citep{2011MNRAS.415.3824S}. 

After the discovery of GW 150914, searches for EM counterparts of stellar-mass BH-BH mergers have become of more interest.  
A number of follow-up EM observations were indeed made from optical to gamma-ray energy bands~\citep{GW15094_PanSTARRS,GW15094_Swift,2016arXiv160204180S,GW15094_LAT}. 
A possible low significance association with a short-duration gamma-ray burst (GRB) event has been reported with Fermi's Gamma-ray Burst Monitor (GBM) just 0.4 s after 
GW 150914~\citep{GW15094_GBM}.  It is tentative and this signal has not been confirmed by INTEGRAL~\citep{2016arXiv160204180S}, but a number of possibilities to account for this have already been advanced~\citep{Loeb_2016,Zhang_2016,Li_et_al_2016,2016arXiv160205140P}.  
A possible EM counterpart due to super-Eddington accretion onto the BH is discussed for BH-BH binaries embedded in active galaxies~\citep{2016arXiv160203831B,2016arXiv160204226S}. 
Afterglow emission of relativistic jets \citep{2016arXiv160205050Y,2016arXiv160205529M} has also been considered. 
Furthermore, the combination of EM and GW signals from the same source has been used to test the Einstein's equivalence principle~\citep{2016arXiv160201566W} and
modified dispersion relations for GWs~\citep{2016arXiv160205882C}. 

In this work, we consider the fate of a possible mini-disk accompanied by a BH binary.  Our aim is to reveal consequences of such a system and to show that a possible EM counterpart signal can be used to test the proposed models.  Our work does not rely on the tentative association with a short GRB.  
In Sec. 2 we consider fast optical transients that can emerge from disk-driven outflows. In Sec. 3 we study the possibility of long-lasting radio emission from blast waves originating from ultrafast outflows. In Sec. 4 we discuss a possible jet component.  Throughout this work, we use the notation $Q={10}^xQ_x$ in CGS unit unless noted otherwise.

\section{Optical transients from a disk wind}
If an accretion disk exists around a BH, a disk wind may be driven by radiation and/or magnetic fields in the disk corona.  
In particular, strong disk winds are commonly suggested in numerical studies of super-Eddington accretion disks
~\citep[e.g.,][]{Ohsuga_et_al_2005,Jiang_et_al_2014,Sadowski_et_al_2014}.
Although the origin of the disk material is an open question, it has been suggested that a low-temperature 
``fossil'' disk may exist around one of the BHs via the formation of the dead zone~\citep{Perna_et_al_2014}.  
The disk may become ionized and active around the coalescence of the BHs~\citep{2016arXiv160205140P}.  
The viscous time for the disk around a BH is 
\begin{equation}
t_{\rm vis}=\frac{1}{\alpha\Omega_K}{\left(\frac{R_d}{H}\right)}^2\simeq1.4~{\rm s}~M_{\rm BH,1.78}^{-1/2}R_{d,8}^{3/2}{\alpha}_{-1}^{-1}{(h/0.3)}^{-2},
\end{equation}
where $\alpha$ is the viscosity parameter, $H$ is the disk scale height, $\Omega_K=\sqrt{GM_{\rm BH}/R_d^3}$ is the Kepler rotation frequency, 
$M_{\rm BH}=60~M_\odot~M_{\rm BH,1.78}$ is the merged BH mass, and $h\equiv H/R_d$. 
Equation~(1) may be applied once the magnetorotational instability (MRI) becomes effective.  
The MRI time scale is $t_{\rm MRI}\approx(4/3\Omega_K)\simeq0.015~{\rm s}~M_{\rm BH,1.78}^{-1/2}R_{d,8}^{3/2}$, which is essentially the dynamical time.
In this work, for illustrative purposes, we take the BH mass to be $M_{\rm BH}\sim10-100~M_\odot$, a disk mass 
$M_{d}\sim{10}^{-5}-1~M_\odot$ (which may be comparable to the Jupiter mass), and a disk size 
$R_d\sim{10}^{8}-{10}^{11}$~cm as reference parameters~\citep[e.g.,][]{2016arXiv160205140P}.  
One can easily consider cases of other BH-disk parameters.
In the fossil disk model, the mass accretion rate is 
$\dot{M}_d\approx M_d/t_{\rm vis}\simeq7.0\times{10}^{-4}~M_{\odot}~{\rm s}^{-1}~M_{d,-3}
M_{\rm BH,1.78}^{1/2}R_{d,8}^{-3/2}{\alpha}_{-1}{(h/0.3)}^{2}$.
Such a super-Eddington accretion event might also happen even {\it before or after} the merger depending on models.  
The post-merger violent accretion occurs if the disk remains cold by the merger time $t_{\rm GW}$ becomes shorter than $t_{\rm vis}$. 
On the other hand, the pre-merger accretion may occur if the disk is ionized for $t_{\rm GW}>t_{\rm vis}$.
Alternatively, for a hierarchical three-body system, a star can be trapped resonantly~\citep{2011MNRAS.415.3824S} and may lead to the tidal disruption.  
We expect that not only stars but also planets can be involved, and planet formation around a BH is also suggested~\citep{Perna_et_al_2014}. 
The tidal radius for a star with $R_*$ and $M_*$ is 
$R_{t}\approx R_*{(M_{\rm BH}/M_*)}^{1/3}\simeq2.5\times{10}^{11}~{\rm cm}~M_{\rm BH,1.78}^{1/3}M_{*,0}^{-1/3}R_{*,10.8}$.  
The return time of the bound material is $t_{\rm td}\approx\pi(R_{*}^3M_{\rm BH}/2GM_*^2)^{1/2}\simeq2.4\times{10}^4~{\rm s}~
M_{\rm BH,1.78}^{1/2}M_{*,0}^{-1}R_{*,10.8}^{3/2}$, leading to 
$M_*/(3t_{\rm td})\simeq1.4\times{10}^{-5}~M_\odot~{\rm s}^{-1}~M_{\rm BH,1.78}^{-1/2}M_{*,0}^{2}R_{*,10.8}^{-3/2}$~\citep{1989ApJ...346L..13E}.

We parametrize the disk outflow rate as ${\dot M}_{w}=\eta_{w}\dot{M}_d$, which is 
${\dot M}_{w}\sim{10}^{-6}-{10}^{-3}~M_\odot~{\rm s}^{-1}$ (i.e. $\eta_w\sim0.1-1$) as our typical parameters.  
The disk wind velocity $v_w$ is expected to be comparable to the escape velocity,
\begin{equation} 
v_{\rm esc}={(2GM_{\rm BH}/R_d)}^{1/2}\simeq0.4~c~M_{\rm BH,1.78}^{1/2}R_{d,8}^{-1/2},
\end{equation}
which can be a significant fraction of the speed of light $c$. Throughout this work, we assume a constant wind velocity (although radiative acceleration is possible). 
The density at the foot-point of the wind is so large that it is expected to be radiation-dominated~\citep[cf.][for studies on tidal disruption events and massive stellar collapses]{Rossi_Begelman_2009,Strubbe_Quataert_2009,Kashiyama_Quataert_2015}. 
Its initial temperature is $T_0\simeq1.3\times{10}^{9}~{\rm K}~{\dot M}_{w,-4}^{1/4}M_{\rm BH,1.78}^{1/8}R_{d,8}^{-5/8}$. 
The optical depth is defined by $\tau_T=\kappa\rho v_wt$ (where $\kappa$ is the opacity). 
The Thomson optical depth at $R_d$ is $\tau_T^0\simeq4.5\times{10}^9~M_{\rm BH,1.78}^{-1/2}R_{d,8}^{-1/2}{\dot M}_{w,-4}(\kappa_T/0.34~{\rm cm}^2~{\rm g}^{-1})$, where $\kappa_T$ is the Thomson scattering opacity.

The disk wind can be regarded as a continuous outflow until 
$r_w\approx v_wt_{\rm acc}$, where $t_{\rm acc}$ is $t_{\rm vis}$ or $t_{\rm td}$.
We have $r_w\simeq1.7\times{10}^{10}~{\rm cm}~R_{d,8}{\alpha}_{-1}^{-1}{(h/0.3)}^{-2}$ 
in the fossil disk model and $r_w\simeq9.1\times{10}^{12}~{\rm cm}~R_{d,11}^{-1/2}M_{\rm BH,1.78}M_{*,0}^{-1}R_{*,10.8}^{3/2}$ in the tidal disruption model, respectively.
The temperature and density scale as $T\propto r^{-2/3}$ and $\rho\propto r^{-2}$, so that 
we expect $\tau_T\propto r^{-1}$ for $r<r_w$. 
The disk wind effectively~\footnote{Note that we expect $\dot{M}_d(t>t_{\rm td}) \propto t^{-\beta}$ and $\beta=5/3$ in the tidal disruption case. } 
ceases at $t_{\rm acc}$, and for $r>r_{w}$ we may expect a  homologous 
expansion of the outflow. Thereafter, the temperature and density scale as $T\propto r^{-1}$ 
and $\rho\propto r^{-3}$, leading to $\tau_T\propto r^{-2}$. The Thomson optical depth at 
$r>r_w$ is $\tau_T\simeq1.4\times{10}^7~M_{\rm BH,1.78}^{-1/2}R_{d,8}^{1/2}{\dot M}_{w,-4}
r_{w,10.5}^{-1}(\kappa_T/0.34~{\rm cm}^2~{\rm g}^{-1}){(r/r_w)}^{-2}$. 

Initially, the photons are trapped in the outflow. But photons start to escape when the 
Thomson optical depth becomes
\begin{equation}
\tau_T^{\rm bo}\approx c/v_w\simeq2.5~M_{\rm BH,1.78}^{-1/2}R_{d,8}^{1/2}.
\end{equation}
The condition $\tau_T(r_{\rm bo})=\tau_T^{\rm bo}$ gives the photon breakout radius
\begin{equation}
r_{\rm bo}\simeq7.5\times{10}^{13}~{\rm cm}~{\dot M}_{w,-4}^{1/2}r_{w,10.5}^{1/2}{(\kappa_T/0.34~{\rm cm}^2~{\rm g}^{-1})}^{1/2}.
\end{equation}
Note that around this radius the flow expansion time is comparable to the photon diffusion 
time.  The diffusion timescale at this radius is estimated to be
\begin{eqnarray}
t_{\rm diff}^{\rm bo}\approx\frac{\tau_T^{\rm bo}r_{\rm bo}}{c}&\simeq&6300~{\rm s}~M_{\rm BH,1.78}^{-1/2}R_{d,8}^{1/2}\nonumber\\
&\times&{\dot M}_{w,-4}^{1/2}r_{w,10.5}^{1/2}{(\kappa_T/0.34~{\rm cm}^2~{\rm g}^{-1})}^{1/2}.
\end{eqnarray}
After $t_{\rm diff}^{\rm bo}$, the thermal photons escape from the outflow, and can
be observed by optical telescopes.  Indeed, the typical temperature of this thermal emission is
\begin{eqnarray}
T_{\rm bo}&\approx&1.1\times{10}^{4}~K~M_{\rm BH,1.78}^{1/8}R_{d,8}^{1/24}\nonumber\\
&\times&{\dot M}_{w,-4}^{-1/4}r_{w,10.5}^{-1/6}{(\kappa_T/0.34~{\rm cm}^2~{\rm g}^{-1})}^{-1/2}.
\end{eqnarray}
Notably, the value is quite insensitive to various parameters such as $M_{\rm BH}$, $R_d$, 
${\dot M}_{w}$, and $r_{w}$. Thus, the predictions about these fast optical transients 
are promising as long as a mini-disk exists in a BH-BH merger. 
The peak bolometric luminosity is estimated to be
\begin{eqnarray}
L_{\rm th}^{\rm pk}\approx\frac{4\pi r_{\rm bo}^3 a T_{\rm bo}^4}{3t_{\rm diff}^{\rm bo}}&\simeq&3.6\times{10}^{40}~{\rm erg}~{\rm s}^{-1}~M_{\rm BH,1.78}R_{d,8}^{-1/3}\nonumber\\
&\times&r_{w,10.5}^{1/3}{(\kappa_T/0.34~{\rm cm}^2~{\rm g}^{-1})}^{-1}.
\end{eqnarray}
For the above nominal values the bolometric flux is $F_{\rm bol}=L_{\rm bol}/(4\pi d^2)\simeq2.9\times10^{-14}~d_{26.5}^{-2}$~erg~cm$^{-2}$~s$^{-1}$
(where $d$ is the distance to the source); 
the spectral peak $\nu_{\rm pk}=2.82k_B T_{\rm bo}/h\simeq6.5\times10^{14}$ Hz is in the B- or V-band; 
the spectral flux $F_\nu\sim4.4\times 10^{-29}$~erg~cm$^{-2}$~s$^{-1}$~Hz$^{-1}=4.4~\mu$Jy corresponds to a magnitude $m\sim22+5\log(d_{26.5})$.
Note also that the bolometric luminosity is proportional to $M_{\rm BH}$.  
Thus, it is easier to see the EM counterparts of BH mergers involving more massive BHs. 
Knowing the detailed shape of the BH mass function in BH-BH mergers, which depends on 
formation scenarios~\citep[e.g.,][]{2016ApJ...818L..22A,2014MNRAS.442.2963K,2016arXiv160204531B,2016arXiv160202809O}, would be relevant to predict the detection rate of the fast optical transients. 
 
Until the outflow reaches the photospheric radius, the bolometric luminosity is roughly 
constant, i.e. $L_{\rm th}^{\rm pk}\propto t^0$.   
The photospheric radius is 
\begin{eqnarray}
r_{\rm ph}&\approx&1.2\times{10}^{14}~{\rm cm}~M_{\rm BH,1.78}^{-1/4}R_{d,8}^{1/4}\nonumber\\
&\times&{\dot M}_{w,-4}^{1/2}r_{w,10.5}^{1/2}{(\kappa_T/0.34~{\rm cm}^2~{\rm g}^{-1})}^{1/2},
\end{eqnarray}
which is reached at the time
\begin{eqnarray}
t_{\rm ph}\approx r_{\rm ph}/v_{w}&\simeq&9900~{\rm s}~M_{\rm BH,1.78}^{-3/4}R_{d,8}^{3/4}\nonumber\\
&\times&{\dot M}_{w,-4}^{1/2}r_{w,10.5}^{1/2}{(\kappa_T/0.34~{\rm cm}^2~{\rm g}^{-1})}^{1/2}.\,\,\,\,\,
\end{eqnarray}
The bolometric luminosity is thereafter expected to rapidly drop as $L_{\rm th}^{\rm pk}
\propto t^{-2}$ just after $t_{\rm ph}$ and shows the exponential decay at $t\gg t_{\rm ph}$.

As shown above, the duration of the expected thermal emission is rather short, lasting 
from hours to days, which also lies at the frontier of optical surveys~\citep{2012arXiv1202.2381K}. 
Typical surveys in the present day (e.g., Pan-STARRS) may achieve a photometric magnitude of $m\sim20-22$, 
which is hard to see the event at $d\sim400$~Mpc but nearby post-merger emission could be seen. 
Future LSST (with $m\sim24.5$) is more promising. 
There may be possible confusions with optical transients from e.g., super-Eddington outbursts of Galactic X-ray binaries~\citep[e.g.,][]{2002A&A...385..904R}, 
but they are expected to show persistent emission that can be distinguished.
Even though it is challenging to make a follow-up observation, detecting such 
short optical transients would enable us to unequivocally identify the EM counterparts of BH-BH mergers. 
Blind searches with optical monitors with a wide field of view would be relevant for testing the model. 
The localization of the BH-BH mergers would in turn allow us to study their host galaxies, environments, and formation mechanisms.

\section{Radio emission}
An ultrafast flow originating from a mini-disk wind develops into a blast wave, which 
starts to slow down at the deceleration radius
\begin{equation}
r_{\rm dec}\approx {\left(\frac{3M_{w}}{4\pi nm_p}\right)}^{1/3}\simeq3.1\times{10}^{17}~{\rm cm}~M_{w,-4}^{1/3}n^{-1/3},
\end{equation}
which is essentially the Sedov radius.  Here $n$ is the ambient density.  The corresponding deceleration time is
\begin{equation}
t_{\rm dec}\approx r_{\rm dec}/v_w\simeq2.5\times{10}^{7}~{\rm s}~M_{\rm BH,1.78}^{-1/2}R_{d,8}^{1/2}M_{w,-4}^{1/3}n^{-1/3}.
\end{equation}
The shock velocity at $t>t_{\rm dec}$ is 
$v\approx0.4v_w{(t/t_{\rm dec})}^{-3/5}\simeq6.1\times{10}^{9}~{\rm cm}~{\rm s}^{-1}~M_{\rm BH,1.78}^{1/5}
R_{d,8}^{-1/5}M_{w,-4}^{1/5}n^{-1/5}t_{7.5}^{-3/5}$.
Assuming an adiabatic index $\hat{\gamma}=5/3$, the post-shock magnetic field is estimated to be
$B={(9\pi\epsilon_Bnm_pv^2)}^{1/2}\simeq4.2~{\rm mG}~M_{\rm BH,1.78}^{1/5}R_{d,8}^{-1/5}M_{w,-4}^{1/5}n^{3/10}\epsilon_{B,-2}^{1/2}t_{7.5}^{-3/5}$,
where $\epsilon_B$ is the energy fraction carried by magnetic fields compared to the 
downstream thermal energy density. We take $\epsilon_B\sim0.01$ as used in the literature 
of gamma-ray bursts and trans-relativistic supernovae~\citep{2006RPPh...69.2259M}. 

We expect that electrons are accelerated at the forward shock, which leads to 
broadband synchrotron emission~\citep[cf.][for GRBs]{1993ApJ...405..278M}. 
The injection Lorentz factor of electrons at a non-relativistic shock is given by
\begin{eqnarray}
\gamma_{ei}&\approx&(\zeta_e/2)(m_p/m_e)(v^2/c^2)\nonumber\\
&\simeq7.2&~M_{\rm BH,1.78}^{2/5}R_{d,8}^{-2/5}M_{w,-4}^{2/5}n^{-2/5}(\zeta_e/0.4)t_{7.5}^{-6/5},\,\,\,
\end{eqnarray}
where $\zeta_e$ is a numerical coefficient related the energy fraction and injection 
fraction of accelerated electrons. We adopt $\zeta_e\sim0.4$ based on the recent results 
of particle-in-cell simulations~\citep{2015PhRvL.114h5003P}. 
The accelerated electrons cool mainly via synchrotron 
radiation, on a time scale $t_{\rm syn}\approx 6\pi m_ec/(\sigma_T B^2\gamma_e)$. From the condition $t_{\rm syn}=t$, 
the cooling Lorentz factor of the electrons is estimated to be
\begin{eqnarray}
\gamma_{ec}&\approx&\frac{6\pi m_ec}{\sigma_TB^2 t}\nonumber\\
&\simeq&2.9\times{10}^6~M_{\rm BH,1.78}^{-2/5}R_{d,8}^{2/5}M_{w,-4}^{-2/5}n^{-3/5}\epsilon_{B,-2}^{-1}t_{7.5}^{1/5}.\,\,\,\,\,
\end{eqnarray}
The acceleration time of electrons via diffusive shock acceleration is given by 
$t_{\rm acc}\approx(20/3)c\gamma_em_ec^2/(eBv^2)$ in the Bohm limit.
The condition $t_{\rm acc}=t_{\rm syn}$ gives the maximum Lorentz factor of electrons, 
which is 
\begin{eqnarray}
\gamma_{eM}\approx{\left(\frac{9\pi ev^2}{10\sigma_TB c^2}\right)}^{1/2}&\simeq&1.2\times{10}^8~M_{\rm BH,1.78}^{1/10}R_{d,8}^{-1/10}\nonumber\\
&\times&M_{w,-4}^{1/10}n^{-7/20}\epsilon_{B,-2}^{-1/4}t_{7.5}^{-3/10}.\,\,\,\,\,\,\,\,\,\,\,\,\,
\end{eqnarray}

With the above parameters synchrotron spectra can now be calculated. Since we have
$\gamma_{ei}\ll\gamma_{ec}$, the resulting spectrum is expected in the slow-cooling regime. 
The injection synchrotron frequency $\nu_i$ is given by
\begin{eqnarray}
\nu_{i}\approx\gamma_{ei}^2\frac{3eB}{4\pi m_ec}&\simeq&6.4\times{10}^5~{\rm Hz}~M_{\rm BH,1.78}R_{d,8}^{-1}\nonumber\\
&\times&M_{w,-4}n^{-1/2}\epsilon_{B,-2}^{1/2}{(\zeta_e/0.4)}^{2}t_{7.5}^{-3},
\end{eqnarray}
the cooling synchrotron frequency $\nu_c$ is
\begin{equation}
\nu_{c}\simeq1.0\times{10}^{17}~{\rm Hz}~M_{\rm BH,1.78}^{-3/5}R_{d,8}^{3/5}M_{w,-4}^{-3/5}n^{-9/10}\epsilon_{B,-2}^{-3/2}t_{7.5}^{-1/5}, 
\end{equation}
and the maximum synchrotron frequency is
\begin{equation}
\nu_{M}\simeq1.7\times{10}^{20}~{\rm Hz}~M_{\rm BH,1.78}^{2/5}R_{d,8}^{-2/5}M_{w,-4}^{2/5}n^{-2/5}t_{7.5}^{-6/5}.
\end{equation}
The peak synchrotron flux, which occurs at $\nu_i$ in the slow-cooling case, is 
\begin{eqnarray}
F_\nu^{\rm max}&\approx&\frac{0.6f_e nr^3}{4\pi d^2}\frac{4\sqrt{3}\pi e^3B}{3m_ec^2}\nonumber\\
&\simeq&5.0~{\rm mJy}~M_{\rm BH,1.78}^{4/5}R_{d,8}^{-4/5}M_{w,-4}^{4/5}\nonumber\\
&\times&n^{7/10}f_e\epsilon_{B,-2}^{1/2}t_{7.5}^{3/5}d_{26.5}^{-2},
\end{eqnarray}
where $f_e$ is the number fraction of accelerated electrons. 
The synchrotron spectrum at $\nu_i<\nu<\nu_c$ is $F_{\nu}\propto \nu^{1/2-s/2}$, where $s$ 
is the injection spectral index of the accelerated electrons (that is defined by $dN_e/d\gamma_e\propto\gamma_e^{-s}$).  
The spectrum becomes $F_{\nu}\propto \nu^{-s/2}$ at $\nu_c<\nu<\nu_M$.  
The radio and optical band typically lies in the range of $\nu_i<\nu<\nu_c$, where the
synchrotron flux at time $t$ is approximately given by
\begin{eqnarray}
F_\nu&\sim&0.03~{\rm mJy}~\nu_9^{1/2-s/2}M_{\rm BH,1.78}^{3/10+s/2}R_{d,8}^{-3/10-s/2}M_{w,-4}^{3/10+s/2}\nonumber\\
&\times&n^{19/10-s/4}f_e\epsilon_{B,-2}^{1/4+s/4}{(\zeta_e/0.4)}^{-1+s}t_{7.5}^{21/10-3s/2}d_{26.5}^{-2}.\,\,\,\,\,\,\,\,\,\,
\end{eqnarray}
The detection of non-thermal radio signals is possible for nearby BH-BH mergers, 
unless the ambient number density is too small.  
For comparison, the Very Large Array has a sensitivity of $\sim0.03-0.1$~mJy.
An advantage of the synchrotron radio signals is that the emission is long-lasting, so that follow-up 
observations can be made on the scale of months-to-years after the detections of GW signals 
by Advanced LIGO, Advanced VIRGO and KAGRA~\citep{2012CQGra..29l4007S}.

\section{Relativistic jets and cosmic rays}
In principle, the post-merger emission from a relativistic jet could also be expected for the same BH-disk system.
In particular, if the association with a short-duration GRB detected by Fermi-GBM 
is real, such a jet component is necessary to explain the observed gamma-ray luminosity
of $L_{\gamma}^{\rm iso}\sim{10}^{49}~{\rm erg}~{\rm s}^{-1}$. 
As commonly discussed in the literature of EM counterparts of supermassive BH binaries~\citep{2011CQGra..28i4021S,2010PhRvD..81f4017M,2010Sci...329..927P}, 
a merged BH is spinning and its rotation energy can be extracted via the Blandford-Znajek (BZ)
process~\citep{Blandford_Znajek_1977}.  The absolute jet luminosity is limited by~
\citep{Blandford_Znajek_1977,2005ApJ...630L...5M,Tchekhovskoy_et_al_2011}
\begin{eqnarray}
L_{\rm BZ}&\approx& \frac{1}{6c}\Omega_H^2 B_p^2 R_H^4\nonumber\\
&\simeq&1.27\times{10}^{47}~{\rm erg}~{\rm s}^{-1}~B_{p,12}^2M_{\rm BH,1.78}^2,
\end{eqnarray}
for the dimensionless Kerr parameter of $a\sim0.7$.  Here $\Omega_H$ is the BH rotation 
frequency, $R_H$ is the horizon radius, and $B_p$ is the magnetic field anchored to the BH. 
The magnetic field would be supplied by the mini-disk via magnetohydrodynamic instabilities 
such as the MRI, although the formation of ordered magnetic fields in 
the BH magnetosphere is uncertain.  
But a rough upper limit on $B_{p}$ can be placed by $B_p^2/(8\pi)\lesssim\dot{M}_dc/(4\pi R_H^2)$, 
which leads to $B_p\lesssim2.1\times{10}^{13}~{\rm G}~{\dot M}_{d,-3}^{1/2}M_{\rm BH,1.78}^{-1}$,
although the maximum value of $B_p$ seems to require extreme conditions.  Note that the 
BZ outflow may be significantly collimated, where the isotropic-equivalent luminosity is enhanced 
by the inverse of the beaming factor $2\theta_j^{-2}$, where $\theta_j$ is the jet 
opening angle. 
In the fossil disk model, $t_{\rm acc}\approx t_{\rm vis}$ is so short that the jet emission can be more luminous 
than Galactic X-ray binaries, even if the jet launching mechanism may be the same.

The jet luminosity is often expressed as $L_{\rm BZ}=\eta_j {\dot M}_d c^2$. For 
${\dot M}_d\sim{10}^{-3}~M_{\odot}~{\rm s}^{-1}$, $\eta_j$ is expected to be 
$\sim{10}^{-4}-{10}^{-3}$ for $B_p\sim{10}^{12}-{10}^{13}~{\rm G}$.  Noting that the 
disk-wind luminosity is $L_w=\eta_w{\dot M}_d v_w^2\sim \eta_w{\dot M}_d c^2$, 
we obtain $L_{\rm BZ}/L_w\sim {10}^{-2}\eta_{j,-3}\eta_{w,-1}^{-1}$. 
If the BZ outflow is collimated, the jet component can overwhelm the total flux received by an
on-axis observer.  Accordingly, the afterglow radio emission would be dominated by the jet 
component~\citep{2016arXiv160205050Y,2016arXiv160205529M}.
However, note that the wind component is still expected to be significant for 
super-Eddington accretion. In particular, the wind emission would be a dominant component 
for off-axis observers, which is relevant in the search for EM counterparts of a bulk of GW sources.

An interesting question of BH-BH mergers is whether they can be potential cosmic-ray 
accelerators and associated neutrino sources. 
The isotropic-equivalent magnetic luminosity that is required to 
accelerate cosmic rays with energy $E$ is~\citep{2000PhST...85..191B,2004Prama..62..483W}
\begin{eqnarray}
L_B^{\rm iso}\gtrsim\frac{1}{2}\frac{E^2}{Z^2e^2}\Gamma^2 c,
\end{eqnarray}
where $L_B^{\rm iso}$ is the isotropic-equivalent magnetic luminosity, $\Gamma$ is the 
Lorentz factor of the acceleration region, and $Z$ is the charge of the cosmic-ray particles. 
Noting that causality implies the condition $\Gamma\theta_j\gtrsim1$, we have
\begin{equation}
E\lesssim{\left(\frac{4Z^2e^2L_{\rm BZ}}{\Gamma^2\theta_j^2c}\right)}^{1/2}\lesssim2.6\times{10}^{22}~{\rm eV}~Z{\dot M}_{d,-3}^{1/2}.
\end{equation}
Although the maximum value indicated above would be too extreme, BH-BH mergers could be potential accelerators of cosmic rays.  
Correspondingly, they could be sources of high-energy neutrinos as well~\citep[cf.][for supermassive BH binaries]{2011arXiv1103.1886T}, 
although all predictions depend on details of the dissipation and emission mechanisms. 
Interestingly, the observed BH merger rate is not far from the short GRB rate. 
However, even if the cosmic-ray energy per BH merger reaches ${\mathcal E}_{\rm cr}^{\rm iso}\sim{10}^{50}~{\rm erg}$, 
the luminosity density ${\mathcal E}_{\rm cr}^{\rm iso}\rho_{\rm dBH}={10}^{42}~{\rm erg}~{\rm Mpc}^{-3}~{\rm yr}^{-1}
~({\mathcal E}_{\rm cr}^{\rm iso}/{10}^{50}~{\rm erg})(\rho_{\rm dBH}/10~{\rm Gpc}^{-3}~{\rm yr}^{-1})$, 
is far below the energy budget of ultra-high-energy cosmic rays, ${10}^{44}~{\rm erg}~{\rm Mpc}^{-3}~{\rm yr}^{-1}$
~\citep[e.g.,][]{2009ApJ...690L..14M}.

\section{Summary and Discussion}
We considered the fate of ultrafast disk winds from possible mini-disks associated with stellar-mass BH-BH mergers.  
We have shown that: 
(1) fast disk winds will lead to fast optical transients that shine in a timescale from hours to days; 
(2) the outflows interacting with the interstellar medium will cause a strong forward shock, 
and synchrotron emission from the accelerated electrons may expected on a timescale of years. 
The identifications of such EM counterparts are of interest independently of whether the possible association with short GRBs is real or not.
While the detections may be challenging, the existing models involving disks can be tested by dedicated follow-up observations and/or monitoring searches.  

We also discussed a possible physical origin for a tentative short GRB as reported by \cite{GW15094_GBM}.  
The jet powered by the BZ process could potentially give a viable explanation for the observed luminosity in terms of a relativistic jet component, if the disk exists. 

\begin{acknowledgements}
We acknowledge support by the Pennsylvania State University (K. M. and I. S.), the NASA Einstein
Fellowship program (K. K.), and NASA NNX13AH50G (P. M. and N.S.).   
\end{acknowledgements}

\bibliography{ref}

\begin{thebibliography}{47}
\expandafter\ifx\csname natexlab\endcsname\relax\def\natexlab#1{#1}\fi

\bibitem[{{Abbott} {et~al.}(2016{\natexlab{a}}){Abbott}, {Abbott}, {Abbott},
  {Abernathy}, {Acernese}, {Ackley}, {Adams}, {Adams}, {Addesso}, {Adhikari},
  \& et~al.}]{2016ApJ...818L..22A}
{Abbott}, B.~P., {Abbott}, R., {Abbott}, T.~D., {et~al.} 2016{\natexlab{a}},
  \apjl, 818, L22

\bibitem[{{Abbott} {et~al.}(2016{\natexlab{b}}){Abbott}, {Abbott}, {Abbott},
  {Abernathy}, {Acernese}, {Ackley}, {Adams}, {Adams}, {Addesso}, {Adhikari},
  \& et~al.}]{GW15094_LIGO_Main}
---. 2016{\natexlab{b}}, Physical Review Letters, 116, 061102

\bibitem[{{Bartos} {et~al.}(2016){Bartos}, {Kocsis}, {Haiman}, \&
  {M{\'a}rka}}]{2016arXiv160203831B}
{Bartos}, I., {Kocsis}, B., {Haiman}, Z., \& {M{\'a}rka}, S. 2016, ArXiv
  e-prints, 1602.03831

\bibitem[{{Baruteau} {et~al.}(2012){Baruteau}, {Ramirez-Ruiz}, \&
  {Masset}}]{2012MNRAS.423L..65B}
{Baruteau}, C., {Ramirez-Ruiz}, E., \& {Masset}, F. 2012, \mnras, 423, L65

\bibitem[{{Belczynski} {et~al.}(2016){Belczynski}, {Holz}, {Bulik}, \&
  {O'Shaughnessy}}]{2016arXiv160204531B}
{Belczynski}, K., {Holz}, D.~E., {Bulik}, T., \& {O'Shaughnessy}, R. 2016,
  ArXiv e-prints, 1602.04531

\bibitem[{{Blandford}(2000)}]{2000PhST...85..191B}
{Blandford}, R.~D. 2000, Physica Scripta Volume T, 85, 191

\bibitem[{{Blandford} \& {Znajek}(1977)}]{Blandford_Znajek_1977}
{Blandford}, R.~D., \& {Znajek}, R.~L. 1977, \mnras, 179, 433

\bibitem[{{Collett} \& {Bacon}(2016)}]{2016arXiv160205882C}
{Collett}, T.~E., \& {Bacon}, D. 2016, ArXiv e-prints, 1602.05882

\bibitem[{{Connaughton} {et~al.}(2016){Connaughton}, {Burns}, {Goldstein},
  {Briggs}, {Zhang}, {Hui}, {Jenke}, {Racusin}, {Wilson-Hodge}, {Bhat},
  {Cleveland}, {Fitzpatrick}, {Giles}, {Gibby}, {Greiner}, {von Kienlin},
  {Kippen}, {McBreen}, {Mailyan}, {Meegan}, {Paciesas}, {Preece}, {Roberts},
  {Sparke}, {Stanbro}, {Toelge}, {Veres}, {Yu}, \& {authors}}]{GW15094_GBM}
{Connaughton}, V., {Burns}, E., {Goldstein}, A., {et~al.} 2016, ArXiv e-prints,
  1602.03920

\bibitem[{{Evans} \& {Kochanek}(1989)}]{1989ApJ...346L..13E}
{Evans}, C.~R., \& {Kochanek}, C.~S. 1989, \apjl, 346, L13

\bibitem[{{Evans} {et~al.}(2016){Evans}, {Kennea}, {Barthelmy}, {Beardmore},
  {Burrows}, {Campana}, {Cenko}, {Gehrels}, {Giommi}, {Gronwall}, {Marshall},
  {Malesani}, {Mingo}, {Nousek}, {O'Brien}, {Osborne}, {Pagani}, {Page},
  {Palmer}, {Perri}, {Racusin}, {Siegel}, {Sbarufatti}, \&
  {Tagliaferri}}]{GW15094_Swift}
{Evans}, P.~A., {Kennea}, J.~A., {Barthelmy}, S.~D., {et~al.} 2016, ArXiv
  e-prints, 1602.03868

\bibitem[{{Fermi-LAT collaboration}(2016)}]{GW15094_LAT}
{Fermi-LAT collaboration}. 2016, ArXiv e-prints, 1602.04488

\bibitem[{{Jiang} {et~al.}(2014){Jiang}, {Stone}, \&
  {Davis}}]{Jiang_et_al_2014}
{Jiang}, Y.-F., {Stone}, J.~M., \& {Davis}, S.~W. 2014, \apj, 796, 106

\bibitem[{{Kashiyama} \& {Quataert}(2015)}]{Kashiyama_Quataert_2015}
{Kashiyama}, K., \& {Quataert}, E. 2015, \mnras, 451, 2656

\bibitem[{{Kinugawa} {et~al.}(2014){Kinugawa}, {Inayoshi}, {Hotokezaka},
  {Nakauchi}, \& {Nakamura}}]{2014MNRAS.442.2963K}
{Kinugawa}, T., {Inayoshi}, K., {Hotokezaka}, K., {Nakauchi}, D., \&
  {Nakamura}, T. 2014, \mnras, 442, 2963

\bibitem[{{Kulkarni}(2012)}]{2012arXiv1202.2381K}
{Kulkarni}, S.~R. 2012, ArXiv e-prints

\bibitem[{{Li} {et~al.}(2016){Li}, {Zhang}, {Yuan}, {Jin}, {Fan}, {Liu}, \&
  {Wei}}]{Li_et_al_2016}
{Li}, X., {Zhang}, F.-W., {Yuan}, Q., {et~al.} 2016, ArXiv e-prints, 1602.04460

\bibitem[{{Loeb}(2016)}]{Loeb_2016}
{Loeb}, A. 2016, ArXiv e-prints, 1602.04735

\bibitem[{{McKinney}(2005)}]{2005ApJ...630L...5M}
{McKinney}, J.~C. 2005, \apjl, 630, L5

\bibitem[{{M{\'e}sz{\'a}ros}(2006)}]{2006RPPh...69.2259M}
{M{\'e}sz{\'a}ros}, P. 2006, Reports on Progress in Physics, 69, 2259

\bibitem[{{M\'esz\'aros} \& {Rees}(1993)}]{1993ApJ...405..278M}
{M\'esz\'aros}, P., \& {Rees}, M.~J. 1993, \apj, 405, 278

\bibitem[{{Morsony} {et~al.}(2016){Morsony}, {Workman}, \&
  {Ryan}}]{2016arXiv160205529M}
{Morsony}, B.~J., {Workman}, J.~C., \& {Ryan}, D.~M. 2016, ArXiv e-prints,
  1602.05529

\bibitem[{{M{\"o}sta} {et~al.}(2010){M{\"o}sta}, {Palenzuela}, {Rezzolla},
  {Lehner}, {Yoshida}, \& {Pollney}}]{2010PhRvD..81f4017M}
{M{\"o}sta}, P., {Palenzuela}, C., {Rezzolla}, L., {et~al.} 2010, \prd, 81,
  064017

\bibitem[{{Murase} \& {Takami}(2009)}]{2009ApJ...690L..14M}
{Murase}, K., \& {Takami}, H. 2009, \apjl, 690, L14

\bibitem[{{Nissanke} {et~al.}(2013){Nissanke}, {Kasliwal}, \&
  {Georgieva}}]{2013ApJ...767..124N}
{Nissanke}, S., {Kasliwal}, M., \& {Georgieva}, A. 2013, \apj, 767, 124

\bibitem[{{Ohsuga} {et~al.}(2005){Ohsuga}, {Mori}, {Nakamoto}, \&
  {Mineshige}}]{Ohsuga_et_al_2005}
{Ohsuga}, K., {Mori}, M., {Nakamoto}, T., \& {Mineshige}, S. 2005, ApJ, 628,
  368

\bibitem[{{O'Leary} {et~al.}(2016){O'Leary}, {Meiron}, \&
  {Kocsis}}]{2016arXiv160202809O}
{O'Leary}, R.~M., {Meiron}, Y., \& {Kocsis}, B. 2016, ArXiv e-prints,
  1602.02809

\bibitem[{{Palenzuela} {et~al.}(2010){Palenzuela}, {Lehner}, \&
  {Liebling}}]{2010Sci...329..927P}
{Palenzuela}, C., {Lehner}, L., \& {Liebling}, S.~L. 2010, Science, 329, 927

\bibitem[{{Park} {et~al.}(2015){Park}, {Caprioli}, \&
  {Spitkovsky}}]{2015PhRvL.114h5003P}
{Park}, J., {Caprioli}, D., \& {Spitkovsky}, A. 2015, Physical Review Letters,
  114, 085003

\bibitem[{{Perna} {et~al.}(2014){Perna}, {Duffell}, {Cantiello}, \&
  {MacFadyen}}]{Perna_et_al_2014}
{Perna}, R., {Duffell}, P., {Cantiello}, M., \& {MacFadyen}, A.~I. 2014, ApJ,
  781, 119

\bibitem[{{Perna} {et~al.}(2016){Perna}, {Lazzati}, \&
  {Giacomazzo}}]{2016arXiv160205140P}
{Perna}, R., {Lazzati}, D., \& {Giacomazzo}, B. 2016, ArXiv e-prints,
  1602.05140

\bibitem[{{Revnivtsev} {et~al.}(2002){Revnivtsev}, {Sunyaev}, {Gilfanov}, \&
  {Churazov}}]{2002A&A...385..904R}
{Revnivtsev}, M., {Sunyaev}, R., {Gilfanov}, M., \& {Churazov}, E. 2002, \aap,
  385, 904

\bibitem[{{Rossi} \& {Begelman}(2009)}]{Rossi_Begelman_2009}
{Rossi}, E.~M., \& {Begelman}, M.~C. 2009, MNRAS, 392, 1451

\bibitem[{{Savchenko} {et~al.}(2016){Savchenko}, {Ferrigno}, {Mereghetti},
  {Natalucci}, {Bazzano}, {Bozzo}, {Courvoisier}, {Brandt}, {Hanlon},
  {Kuulkers}, {Laurent}, {Lebrun}, {Roques}, {Ubertini}, \&
  {Weidenspointner}}]{2016arXiv160204180S}
{Savchenko}, V., {Ferrigno}, C., {Mereghetti}, S., {et~al.} 2016, ArXiv
  e-prints, 1602.04180

\bibitem[{{S{\c a}dowski} {et~al.}(2014){S{\c a}dowski}, {Narayan}, {McKinney},
  \& {Tchekhovskoy}}]{Sadowski_et_al_2014}
{S{\c a}dowski}, A., {Narayan}, R., {McKinney}, J.~C., \& {Tchekhovskoy}, A.
  2014, \mnras, 439, 503

\bibitem[{{Schnittman}(2011)}]{2011CQGra..28i4021S}
{Schnittman}, J.~D. 2011, Classical and Quantum Gravity, 28, 094021

\bibitem[{{Seto} \& {Muto}(2011)}]{2011MNRAS.415.3824S}
{Seto}, N., \& {Muto}, T. 2011, \mnras, 415, 3824

\bibitem[{{Smartt} {et~al.}(2016){Smartt}, {Chambers}, {Smith}, {Huber},
  {Young}, {Cappellaro}, {Wright}, {Coughlin}, {Schultz}, {Denneau},
  {Flewelling}, {Heinze}, {Magnier}, {Primak}, {Rest}, {Sherstyuk}, {Stalder},
  {Stubbs}, {Tonry}, {Waters}, {Willman}, {Anderson}, {Baltay}, {Botticella},
  {Campbell}, {Dennefeld}, {Chen}, {Della Valle}, {Elias-Rosa}, {Fraser},
  {Inserra}, {Kankare}, {Kotak}, {Kupfer}, {Harmanen}, {Galbany}, {Gal-Yam},
  {Guillou}, {Lyman}, {Maguire}, {Mitra}, {Nicholl}, {Olivares E},
  {Rabinowitz}, {Razza}, {Sollerman}, {Smith}, {Terreran}, \&
  {Valenti}}]{GW15094_PanSTARRS}
{Smartt}, S.~J., {Chambers}, K.~C., {Smith}, K.~W., {et~al.} 2016, ArXiv
  e-prints, 1602.04156

\bibitem[{{Somiya}(2012)}]{2012CQGra..29l4007S}
{Somiya}, K. 2012, Classical and Quantum Gravity, 29, 124007

\bibitem[{{Stone} {et~al.}(2016){Stone}, {Metzger}, \&
  {Haiman}}]{2016arXiv160204226S}
{Stone}, N.~C., {Metzger}, B.~D., \& {Haiman}, Z. 2016, ArXiv e-prints,
  1602.04226

\bibitem[{{Strubbe} \& {Quataert}(2009)}]{Strubbe_Quataert_2009}
{Strubbe}, L.~E., \& {Quataert}, E. 2009, MNRAS, 400, 2070

\bibitem[{{Tchekhovskoy} {et~al.}(2011){Tchekhovskoy}, {Narayan}, \&
  {McKinney}}]{Tchekhovskoy_et_al_2011}
{Tchekhovskoy}, A., {Narayan}, R., \& {McKinney}, J.~C. 2011, \mnras, 418, L79

\bibitem[{{Thompson} \& {Lacki}(2011)}]{2011arXiv1103.1886T}
{Thompson}, T.~A., \& {Lacki}, B.~C. 2011, ArXiv e-prints, 1103.1886

\bibitem[{{Waxman}(2004)}]{2004Prama..62..483W}
{Waxman}, E. 2004, Pramana, 62, 483

\bibitem[{{Wu} {et~al.}(2016){Wu}, {Gao}, {Wei}, {Fan}, {M{\'e}sz{\'a}ros},
  {Zhang}, {Dai}, {Zhang}, \& {Zhu}}]{2016arXiv160201566W}
{Wu}, X.-F., {Gao}, H., {Wei}, J.-J., {et~al.} 2016, ArXiv e-prints, 1602.1566

\bibitem[{{Yamazaki} {et~al.}(2016){Yamazaki}, {Asano}, \&
  {Ohira}}]{2016arXiv160205050Y}
{Yamazaki}, R., {Asano}, K., \& {Ohira}, Y. 2016, ArXiv e-prints, 1602.05050

\bibitem[{{Zhang}(2016)}]{Zhang_2016}
{Zhang}, B. 2016, ArXiv e-prints, 1602.04542

\end{thebibliography}

\end{document}